\documentclass[11pt]{article}
\usepackage{amsmath}
\usepackage{amssymb}
\usepackage{portland}
\usepackage{rotating}
\newlength{\mytopmargin}
\newlength{\myleftmargin}
\newcommand{\PV}{${\rm P}_{\rm V}\;$}
\newcommand{\PVI}{${\rm P}_{\rm VI}\;$}
\newcommand{\PII}{${\rm P}_{\rm II}\;$}
\newcommand{\PIV}{${\rm P}_{\rm V}\;$}
\newcommand{\PIII}{${\rm P}_{\rm III}\;$}
\newcommand{\PIIIprime}{${\rm P}_{\rm III^{\prime}}\;$}
\newcommand{\threehalf}{
        {\lower0.00ex\hbox{\raise.6ex\hbox{\the\scriptfont0 3}
                           \kern-.5em\slash\kern-.1em\lower.45ex
                                     \hbox{\the\scriptfont0 2}}}}
\newcommand{\half}{
        {\lower0.00ex\hbox{\raise.6ex\hbox{\the\scriptfont0 1}
                           \kern-.5em\slash\kern-.1em\lower.45ex
                                     \hbox{\the\scriptfont0 2}}}}
\newcommand{\quarter}{
        {\lower0.00ex\hbox{\raise.6ex\hbox{\the\scriptfont0 1}
                           \kern-.5em\slash\kern-.1em\lower.45ex
                                     \hbox{\the\scriptfont0 4}}}}
\newcommand{\eighth}{
        {\lower0.00ex\hbox{\raise.6ex\hbox{\the\scriptfont0 1}
                           \kern-.5em\slash\kern-.1em\lower.45ex
                                     \hbox{\the\scriptfont0 8}}}}
\newcommand{\othird}{
        {\lower0.00ex\hbox{\raise.6ex\hbox{\the\scriptfont0 1}
                           \kern-.5em\slash\kern-.1em\lower.45ex
                                     \hbox{\the\scriptfont0 3}}}}

\begin{document}
\vspace{4cm}
\noindent
{\bf $\tau$-FUNCTION EVALUATION OF GAP PROBABILITIES IN ORTHOGONAL \\ 
AND SYMPLECTIC MATRIX ENSEMBLES}

\vspace{5mm}
\noindent
P.J.~Forrester and N.S.~Witte${}^\dagger$

\noindent
Department of Mathematics and Statistics
${}^\dagger$(and School of Physics),
University of Melbourne, \\ 
Victoria 3010, Australia; 
Email: p.forrester@ms.unimelb.edu.au; n.witte@ms.unimelb.edu.au

\small
\begin{quote}
It has recently been emphasized that all known exact evaluations of gap
probabilities for classical unitary matrix ensembles are in fact
$\tau$-functions for certain Painlev\'e systems. We show that all  exact 
evaluations of gap probabilities for classical orthogonal matrix ensembles,
either known or derivable from the existing literature,
are likewise $\tau$-functions for certain Painlev\'e systems. In the case
of symplectic matrix ensembles all  exact evaluations, either
known or derivable from the existing literature, are identified
as the mean of two $\tau$-functions, both of which correspond to
Hamiltonians satisfying the same
differential equation, differing only in the boundary condition.
Furthermore the product of these two $\tau$-functions gives the gap
probability in the corresponding unitary symmetry case, while one of those
$\tau$-functions is the gap probability in the corresponding orthogonal
symmetry case.
\end{quote}

\section{Introduction}
An ensemble of $ N\times N$ random matrices $ X $ with joint probability 
density of the matrix elements proportional to
\begin{equation}\label{1m}
\exp \Big ( \sum_{j=1}^\infty a_j {\rm Tr}(X^j) \Big ) =:
\prod_{j=1}^N g(x_j),
\end{equation}
$x_j$ denoting the eigenvalues,
is invariant under similarity transforms $X \mapsto A^{-1} X A$. In
particular, if $X$ is an Hermitian matrix with real, complex and 
quaternion real elements, labelled by the parameter $\beta$ taking the
values $\beta=1,2$ and 4 respectively, then the subgroups of unitary matrices 
which conserve this feature of $X$ under similarity transformations are
the orthogonal ($\beta = 1$), unitary ($\beta = 2$) and 
unitary symplectic matrices ($\beta = 4$). For this reason the ensemble
is said to have an orthogonal ($\beta = 1$), unitary ($\beta = 2$) or
symplectic symmetry ($\beta = 4$). The eigenvalue probability density
function (PDF) for these ensembles has the explicit form
\begin{equation}\label{2.12}
{1 \over C}
\prod_{j=1}^N g(x_j) \prod_{1 \le j <k \le N} |x_k - x_j|^\beta,
\end{equation}

The function $g(x)$ in (\ref{1m}) and (\ref{2.12}) is referred to as a
weight function. In the cases $\beta = 1$ and $\beta = 2$ the weight
functions
\begin{equation}\label{27a}
g(x) = g_1(x) = \left \{
\begin{array}{ll} e^{-x^2/2}, & {\rm Gaussian} \\
x^{(a-1)/2} e^{-x/2} \: \: (x>0), & {\rm Laguerre} \\
(1-x)^{(a-1)/2}(1+x)^{(b-1)/2} \: \: (-1 < x < 1), & {\rm Jacobi} \\
(1+x^2)^{-(\alpha + 1)/2}, & {\rm Cauchy} \end{array} \right.
\end{equation}
and
\begin{equation}\label{27b}
g(x) = g_2(x) = \left \{
\begin{array}{ll} e^{-x^2}, & {\rm Gaussian} \\
x^{a} e^{-x} \: \: (x>0), & {\rm Laguerre} \\
(1-x)^{a}(1+x)^{b} \: \: (-1 < x < 1), & {\rm Jacobi} \\
(1+x^2)^{-\alpha}, & {\rm Cauchy} \end{array} \right.
\end{equation}
are said to define classical matrix ensembles with an orthogonal and
unitary symmetry respectively, or simply classical orthogonal and unitary
ensembles
(a similar definition applies in the symplectic case --- see
e.g.~\cite{AFNV99}).
We recall (see e.g.~the introduction of \cite{FR99})
that the Cauchy ensemble includes
as a special case the PDF
\begin{equation}\label{1.6}
{1 \over C} \prod_{1 \le j < k \le N}
| e^{i \theta_k} - e^{i \theta_j} |^\beta,
\end{equation}
which specifies the Dyson circular ensembles. The two ensembles are related
by the 
stereographic projection
\begin{equation}\label{1.6a}
e^{i \theta_j} = {1 - i x_j \over 1 + i x_j}.
\end{equation}
In particular, changing variables in (\ref{1.6}) according to (\ref{1.6a})
gives a PDF of the form (\ref{2.12}) with $g(x)$ a Cauchy weight function,
which in the cases
$\beta=1$ and 2 is specified by (\ref{27a}) and (\ref{27b}) with
$\alpha = N$.

Our interest is in a special property of the probability
$E_\beta(0;I;g(x);N)$ of having no eigenvalues in the interval $I$
when the eigenvalue PDF is specified by (\ref{2.12}) in the case that 
$g(x)$ is classical. The probability is specified as a multiple integral by
\begin{equation}
  E_\beta(0;I;g(x);N) = 
  {1 \over C} \prod_{j=1}^N \int_{I_0\backslash I} dx_j \, g(x_j) 
   \prod_{1 \le j < k \le N} |x_k - x_j|^\beta,
\end{equation}
where $I_0$ is the interval of support of $g(x)$.
The special property is that for $g(x)$ classical $E_\beta$ admits
Painlev\'e transcendent evaluations
for certain $I$ (the evaluations are in some cases restricted also to
certain scaled limits). We will focus on a
structural aspect of these formulas, by showing that in the orthogonal 
case all known Painlev\'e transcendent evaluations can be identified as
$\tau$-functions for Hamiltonians associated with the
Painlev\'e functions, and  in the symplectic case as the mean of
two $\tau$-functions.

Our work builds on the recently
emphasized \cite{FW00,BD00} fact
that all gap probabilities for 
classical unitary ensembles that have been
characterized as the solution of a single differential equation,
are in fact $\tau$-functions for certain Painlev\'e systems.
Such characterizations of the gap probability for classical unitary matrix
ensembles are known when  the gap consists of a single
interval including an end-point of the support
\cite{TW94c,ASV95,HS99,WF99b,BD00}, or a double
interval symmetrically placed about the origin again including the
end-points of the support or the origin (applicable to even weight
functions only) \cite{TW94c,WFC00,WF99b}.
In
the special case of the gap probability for scaled, infinite GUE matrices in
the bulk, the identification as a $\tau$-function for a
Painlev\'e system was made by Okamoto and quoted in the
original paper of Jimbo et al.~\cite[pg.~152]{JMMS80} deriving the Painlev\'e
evaluation. For the more general problem of characterizing the
gap probabilities in the case of multiple excluded intervals, the
fact that the probability is the
$\tau$-function for certain
integrable systems associated with monodromy preserving deformations
of linear differential equations with rational coefficients was
a main theme of \cite{JMMS80}, and then generalized to a more
general setting (but not the most general case of interest in
random matrix theory) by Palmer \cite{Pa94}. 
Harnad and Its \cite{HI01} have recently discussed the work of
Palmer from a Riemann-Hilbert problem perspective. 
Identifications of the
gap probabilities in the case of multiple excluded intervals
as $\tau$-functions in the Sato
theory is a theme of the work of Adler, van Moerbeke and collaborators
(see e.g.~\cite{AV95}).

The situation with the exact evaluation of gap probabilities for
matrix ensembles  with an orthogonal symmetry is immediately
different due to the restricted number of evaluations in terms of
Painlev\'e transcendents presently known  \cite{TW96,Fo99a,Fo99b}.
In the orthogonal case, the exact evaluations can be catalogued into
two distinct mathematical structures --- the finite $N$ ensembles and
their scaling limit for which the $\tau$-function identification is
immediate, and the infinite Gaussian and Laguerre ensembles scaled at
the soft and hard edges respectively in which the known
Painlev\'e transcendent evaluations  reduce to a
$\tau$-function after some calculation. In the symplectic case all
known exact evaluations result from a formula relating the gap
probability in the symplectic case to that in the orthogonal and
unitary cases. Further special features of the exact evaluations
in the orthogonal and
unitary cases then allows the exact evaluations in the symplectic case
to be identified as the mean of two $\tau$-functions, both of which
correspond to Hamiltonians satisfying the same differential equation, 
differing only in the boundary
condition.

\section{Orthogonal matrix ensembles}
\setcounter{equation}{0}
\subsection{Finite $N$ ensembles}
It has been shown in \cite{Fo99a} that for the classical weights
(\ref{27a}), having the additional property of being even
(which is the case for the Gaussian, symmetric Jacobi $(a=b)$
and Cauchy weights),
\begin{equation}\label{F1}
E_1(0;(-s,s);g_1(x);N) =
E_2(0;(0,s^2); x^{-1/2} g_2(x^{1/2}); N/2)
\end{equation}
where on the RHS $x>0$, and it is assumed $N$ is even.
Now a unitary ensemble with weight
$x^{-1/2} g_2(x^{1/2})$, in which $g_2(x)$ is an even classical weight,
is equal to another unitary ensemble with a classical
weight, after a suitable change of variables as detailed in Table
\ref{t1}.
Hence it follows that
\begin{multline}\label{F2}
  E_2(0;(0,s^2); x^{-1/2} g_2(x^{1/2}); N/2) \\
  = \left\{ 
    \begin{array}{ll}
        E_2(0;(0,s^2); x^{-1/2} e^{-x}; N/2),
      & \text{Gaussian} \\
        E_2(0;(-1,2s^2-1); (1+x)^{-1/2}(1-x)^a; N/2),
      & \text{symmetric Jacobi} \\
        E_2(0;(-1,(s^2-1)/(s^2+1));(1+x)^{-1/2}(1-x)^{\alpha-N+1/2}; N/2),
      & \text{Cauchy}
    \end{array} \right.
\end{multline}
This substituted in (\ref{F1}) gives  $E_1(0;(-s,s);g_1(x);N)$
for the even classical
orthogonal ensembles in terms of $E_2$ for certain classical unitary
ensembles. The latter furthermore
have the gap free interval including
an end-point of the support of the weight function. In such a  case,
we can deduce from the existing literature that $E_2$, and
consequently $E_1$, is a $\tau$-function for an appropriate
Painlev\'e system.

\begin{table}
\begin{center}
\begin{tabular}{|c| c |c|} \hline
$g_2(x)$ & $x^{-1/2} g_2(x^{1/2})$ & classical ensemble \\ \hline
$e^{-x^2}$ & $x^{-1/2} e^{-x}$ & Laguerre: $a=-1/2$ \\
$(1-x^2)^a$ & $x^{-1/2} (1 - x)^a$
 & Jacobi: $x \mapsto 2x - 1$, $b=-{1 \over 2}$
\\
$(1+x^2)^{-\alpha}$ & $x^{-1/2}(1+x)^{-\alpha} $ &
Jacobi: $x \mapsto {x - 1 \over x + 1}$, $b=-1/2$, $a=\alpha - N
+ {1 \over 2}$ \\ \hline
\end{tabular}
\end{center}
\caption{\label{t1} Even classical weights, their transformed form, and
the corresponding classical unitary ensemble with $N/2$ eigenvalues.}
\end{table}

Consider first $E_2(0;(0,s);x^a e^{-x};N)$, specifying the probability
that there are no eigenvalues in the interval $(0,s)$ of the Laguerre
unitary ensemble. Following \cite{Ok87} and \cite{JM81} introduce
the Hamiltonian $H_{\rm V}$ associated with the Painlev\'e V equation by
\begin{equation}\label{6.1}
tH_{\rm V} = q(q-1)^2p^2 - \Big \{ (v_2 - v_1)(q-1)^2 - 2(v_1+v_2)q(q-1)
+tq \Big \}p + (v_3 - v_1)(v_4 - v_1)(q-1)
\end{equation}
where the parameters $v_1,\dots,v_4$ are constrained by
\begin{equation}\label{2.8'}
v_1 + v_2 + v_3 + v_4 = 0.
\end{equation}
The relationship of (\ref{6.1}) to \PV can be seen by eliminating $p$
in the Hamilton equations
\begin{equation}\label{6.1b}
q' = {\partial H \over \partial p}, \qquad
p' = - {\partial H \over \partial q}.
\end{equation}
One finds that $q$ satisfies the equation
\begin{equation}\label{6.1c}
y'' = \Big ( {1 \over 2y} + {1 \over y - 1} \Big ) (y')^2 - {1 \over t} y' 
    + {(y-1)^2 \over t^2} \Big( \alpha y + {\beta \over y} \Big)
    + \gamma {y \over t} + \delta {y(y+1) \over y - 1}
\end{equation}
with
\begin{equation}\label{2.10'}
\alpha = {1 \over 2}(v_3 - v_4)^2, \quad
\beta = - {1 \over 2} (v_2 - v_1)^2, \quad
\gamma = 2 v_1 + 2v_2 - 1, \quad
\delta = - {1 \over 2}.
\end{equation}
This is the general \PV equation with $\delta = - {1 \over 2}$ (recall that
the general \PV equation with $\delta \ne 0$ can be reduced to the case
with $\delta = - {1 \over 2}$ by the mapping $t \mapsto \sqrt{-2 \delta} t$).
Now introduce the auxiliary Hamiltonian
\begin{equation}\label{2.10a}
\sigma_{\rm V} = tH_{\rm V} + (v_3 - v_1)(v_4 - v_1).
\end{equation}
Of course with $tH_{\rm V}$ replaced by $\sigma_{\rm V}$ in  (\ref{6.1b}), the
Hamilton equations remain unchanged so $\sigma_{\rm V}/t$ is also a
Hamiltonian for the same \PV system. The quantity $\sigma_{\rm V}$ satisfies the 
second order, second degree differential equation
\begin{equation}\label{2.g}
  (t \sigma'')^2
   - \left[ \sigma - t \sigma' + 2(\sigma')^2
             + (\nu_0 + \nu_1 + \nu_2 + \nu_3) \sigma' \right]^2
   + 4(\nu_0 + \sigma')(\nu_1 + \sigma')(\nu_2 + \sigma')(\nu_3 + \sigma') = 0,
\end{equation}
with
\begin{equation}\label{2.11a}
\nu_0 = 0, \quad \nu_1 = v_2 - v_1, \quad \nu_2 = v_3 - v_1, \quad
\nu_3 = v_4 - v_1
\end{equation}
(because (\ref{2.g}) is symmetrical in $\{\nu_k\}$ any permutation of
these values is also valid). Conversely, each solution
(with $\sigma'' \ne 0$) of (\ref{2.g}) leads to a solution of the system
(\ref{6.1b}) \cite{Ok87}.

The $\tau$-function associated with the Hamiltonian $\sigma_{\rm V}/t$ is
specified by
\begin{equation}\label{5a}
\sigma_{\rm V} = t {d \over dt} \log \tau_{\sigma_{\rm V}}(t).
\end{equation}
But from the work of Tracy and Widom \cite{TW94c}
we know that
\begin{equation}\label{5b}
t {d \over dt} \log E_2(0;(0,t); x^a e^{-x};N)
\end{equation}
satisfies (\ref{2.g})  with
\begin{equation*}
\nu_0 = 0, \quad \nu_1=0, \quad \nu_2 = N+a, \quad \nu_3 = N,
\end{equation*}
subject to the boundary condition
\begin{equation*}
\sigma(t) \mathop{\sim}\limits_{t \to 0^+} 
{\Gamma(N+a+1) \over \Gamma(N) \Gamma(a+1) \Gamma(a+2)} t^{a+1}.
\end{equation*}
Consequently, after equating (\ref{5a}) and (\ref{5b}), and
normalizing $ \tau_{\sigma_{\rm V}}$ so that $ \tau_{\sigma_{\rm V}}(0)=1$, we have
\begin{equation}\label{5c}
E_2(0;(0,t); x^a e^{-x};N) = \tau_{\sigma_{\rm V}}(t)
\Big |_{\nu_0=0, \: \nu_1=0 \atop \nu_2=N+a, \: \nu_3  = N}.
\end{equation}
With $a=-1/2$ we see that this corresponds to the Gaussian case of
(\ref{F2}). Recalling (\ref{F1}) then gives the sought
 $\tau$-function formula for the Gaussian orthogonal ensemble,
\begin{equation}\label{5d}
E_1(0;(-s,s); e^{-x^2/2};N) = \tau_{\sigma_{\rm V}}(s^2) 
\Big |_{\nu_0=0, \: \nu_1=0 \atop \nu_2=N-1/2, \: \nu_3  = N}.
\end{equation}
 
Consider next $E_2(0;(-1,s);(1-x)^a(1+x)^b;N)$, specifying the
probability that there are no eigenvalues in the interval $(-1,s)$
of the Jacobi unitary ensemble. 
According to (\ref{F2}) this is relevant to both the symmetric Jacobi and
Cauchy cases.
Introduce the Hamiltonian $H_{\rm VI}$
associated with the Painlev\'e VI equation by \cite{Ok87b}
\begin{equation}
t(t-1) H_{\rm VI}
   = q (q-1) (q-t) p^2
     - \left\{ \chi_0(q-1)(q-t) + \chi_1q(q-t) + (\theta - 1) q(q-1)
       \right\} p + \chi (q-t) \label{H6}
\end{equation}
where
\begin{equation*}
\chi = {1 \over 4} ( \chi_0 + \chi_1 + \theta - 1)^2
- {1 \over 4} \chi_\infty^2.
\end{equation*}
Eliminating $p$ from the corresponding Hamilton equations (\ref{6.1b})
shows that $q$ satisfies the \PVI equation
\begin{align*}
y'' = & {1 \over 2} 
      \Big({1 \over y} + {1 \over y - 1} + {1 \over y - t} \Big) (y')^2
      - \Big( {1 \over t} + {1 \over t-1} + {1 \over y - t} \Big) y' \\
      & \qquad
      + {y(y-1)(y-t) \over t^2 (t-1)^2}
        \Big( \alpha + \beta{t \over y^2} + \gamma{t-1 \over (y-1)^2}
              + \delta{t(t-1) \over (y-t)^2} \Big)
\end{align*}
with
\begin{equation*}
\alpha = {1 \over 2} \chi_\infty^2, \quad
\beta = - {1 \over 2} \chi_0^2, \quad
\gamma = {1 \over 2} \chi_1^2, \quad
\delta = {1 \over 2} (1 - \theta^2).
\end{equation*}
Furthermore, the auxiliary Hamiltonian
\begin{equation}\label{h6'}
h_{\rm VI} = t(t-1) H_{\rm VI} + (b_1 b_3 + b_1 b_4 + b_3 b_4 )t
- {1 \over 2} \sum_{1 \le j < k \le 4} b_j b_k,
\end{equation}
where
\begin{equation*}
b_1 = {1 \over 2}(\chi_0 + \chi_1), \quad
b_2 = {1 \over 2}(\chi_0 -  \chi_1), \quad
b_3 = {1 \over 2}(\theta - 1 + \chi_{\infty}), \quad
b_4 = {1 \over 2}(\theta - 1 -  \chi_{\infty}),
\end{equation*}
satisfies the differential equation
\begin{equation}\label{h6}
  h_{\rm VI}' \left( t(1-t) h_{\rm VI}'' \right)^2
  + \left\{ h_{\rm VI}'\left[ 2h_{\rm VI} - (2t-1)h'_{\rm VI} \right]
  + b_1 b_2 b_3 b_4 \right\}^2 = \prod_{k=1}^4(h'_{\rm VI} + b_k^2)
\end{equation}
and conversely, each solution of (\ref{h6}) such that $h''_{\rm VI} \ne 0$ leads
to a solution of the corresponding Hamilton equations.
Now, we know from the work of Haine and Semengue \cite{HS99}, and
Borodin and Deift \cite{BD00}, that
\begin{equation*}
\sigma(t) := t(t-1) {d \over dt} \log
E_2(0;(-1,-1+2t);(1-x)^a(1+x)^b;N) - b_1 b_2 t + {1 \over 2}
(b_1 b_2 + b_3 b_4)
\end{equation*}
with 
\begin{equation*}
b_1 = b_2 = N + {a+b \over 2}, \quad
b_3 = {a+b \over 2}, \quad
b_4 = {a-b \over 2}
\end{equation*}
satisfies (\ref{h6}). Comparing with (\ref{h6'}) we see that with
this choice of parameters
\begin{equation}\label{h7}
 t(t-1) {d \over dt} \log
E_2(0;(-1,-1+2t);(1-x)^a(1+x)^b;N) =
h_{\rm VI} + b_1 b_2 t - {1 \over 2}(b_1 b_2 + b_3 b_4).
\end{equation}
Thus, denoting the RHS of (\ref{h7}) by $\tilde{h}_{VI}$, we see from
(\ref{h6'}) that $\tilde{h}_{\rm VI}/t(t-1)$ is a Hamiltonian for the \PVI
system, and defining the corresponding $\tau$-function by
\begin{equation*}
\tilde{h}_{\rm VI} = t(t-1) {d \over dt} \log \tau_{\tilde{h}_{\rm VI}}(t)
\end{equation*}
we have that
\begin{equation}\label{2.sfj}
E_2(0;(-1,-1+2t);(1-x)^a(1+x)^b;N) =  \tau_{\tilde{h}_{\rm VI}}(t)
\Big |_{b_1 = b_2 = N + (a+b)/2 \atop b_3=(a+b)/2, \: b_4=(a-b)/2}.
\end{equation}
Recalling (\ref{F2}) and (\ref{F1}) then gives the sought $\tau$-function
formulas for the gap probabilities in the Jacobi orthogonal  and
Cauchy orthogonal ensembles,
\begin{align}
  E_1(0;(-s,s);(1-x^2)^{(a-1)/2};N) =
  & \tau_{\tilde{h}_{\rm VI}}(s^2)
    \Big|_{b_1 = b_2 = N/2 + (a-1/2)/2 \atop
           b_3 = (a-1/2)/2, \: b_4 = (a+1/2)/2} \label{au1} \\
  E_1(0;(-s,s);(1+x^2)^{-(\alpha+1)/2};N) =
  & \tau_{\tilde{h}_{\rm VI}} \Big ( {s^2 \over s^2 + 1} \Big )
  \Big |_{b_1 = b_2 = \alpha/2 \atop
          b_3 = (\alpha - N)/2, \: b_4 = (\alpha - N + 1)/2}. \label{au2}
\end{align}

\subsection{Bulk scaling limit}
Let us consider now the $N \to \infty$ bulk scaling limit of an
orthogonal ensemble, and the quantity $E_1^{\rm bulk}(0;2s)$ specifying
the probability that there are no eigenvalues in an interval of length
$2s$ with the mean spacing between eigenvalues equal to unity.
By an appropriate scaling, each of the probabilities in
(\ref{5d}), (\ref{au1}) and (\ref{au2}) tends to $E_1^{\rm bulk}(0;2s)$.
For example, in the Gaussian case the required scaling is
$s \mapsto \pi s/ \sqrt{2N}$ and so
\begin{equation*}
E_1^{\rm bulk}(0;2s) = \lim_{N \to \infty}
E_1(0;(- {\pi s \over \sqrt{2N}}, {\pi s \over \sqrt{2N}});
e^{-x^2/2};N).
\end{equation*}
This scaling applied to (\ref{5d}) is known to lead to the result
\cite{Fo99a}
\begin{equation}\label{csp}
E_1^{\rm bulk}(0;2s) = \exp \Big ( - \int_0^{\pi^2 s^2}
\sigma_B(t) \Big |_{a = -1/2}  \, {dt \over t} \Big )
\end{equation}
where $\sigma_B(t)$ satisfies the equation
\begin{equation}\label{csp1}
(t \sigma_B'')^2 + \sigma_B'(\sigma_B - t\sigma_B')(4\sigma_B' - 1)
                 - a^2 (\sigma_B')^2 = 0
\end{equation}
subject to the boundary condition
\begin{equation}\label{csp2}
  \sigma_B(t)
  \mathop{\sim}\limits_{t \to 0^+} \frac{1}{4}t 
          \left[ J^2_a(\sqrt{t}) - J_{a+1}(\sqrt{t})J_{a-1}(\sqrt{t}) \right]
  \mathop{\sim}\limits_{t \to 0^+}
  {t^{1+a} \over 2^{2+2a} \Gamma(1+a) \Gamma(2+a)}.
\end{equation}

In fact the expression (\ref{csp}) is precisely the $\tau$-function for a
particular \PIII system.
To see this, following Okamoto \cite{Ok87a}, introduce the Hamiltonian
\begin{equation*}
t H = q^2 p^2 - (q^2 + v_1 q - t) p + {1 \over 2} (v_1 + v_2) q.
\end{equation*}
Substituting this form of $H$ in the Hamilton equations (\ref{6.1b})
and eliminating $p$ shows that $y(s) = q(t)/s$, $t=s^2$, satisfies the
general Painlev\'e III equation (Painlev\'e III${}^{\prime}$ in the 
notation of \cite{Ok87a})
\begin{equation*}
  {d^2 y \over ds^2} = {1 \over y} \Big( {dy \over ds} \Big)^2
  - {1 \over s}{dy \over ds} + {1 \over s}(\alpha y^2 + \beta)
  + \gamma y^3 + {\delta \over y}
\end{equation*}
with
\begin{equation*}
\alpha = - 4 v_2, \quad \beta = 4(v_1 + 1), \quad \gamma = 4, \quad
\delta = -4.
\end{equation*}
It is shown in \cite{Ok87a} that the auxiliary quantity
\begin{equation}\label{H4}
h = tH + {1 \over 4} v_1^2 -  {1 \over 2} t
\end{equation} 
satisfies the equation
\begin{equation}\label{H5}
(th'')^2 + v_1 v_2 h' - (4 (h')^2 - 1) (h - th') -
{1 \over 4} (v_1^2 + v_2^2) = 0,
\end{equation}
and conversely all solutions of this equation (assuming $h'' \ne 0$) lead
to the \PIIIprime system. It is a simple exercise to verify from 
the fact that $h$ satisfies (\ref{H5}), the result that 
\begin{equation}\label{cs}
\sigma_{\rm III'}(t) := - (tH) \Big |_{t \mapsto t/4} 
- {v_1 \over 4} (v_1 - v_2) + {t \over 4}
\end{equation}
satisfies the equation
\begin{equation}\label{cs1}
(t \sigma_{\rm III'}'')^2 - v_1 v_2 ( \sigma_{\rm III'}')^2 +
 \sigma_{\rm III'}' (4 \sigma_{\rm III'}' - 1) ( \sigma_{\rm III'} - t  \sigma_{\rm III'}')
- {1 \over 4^3} (v_1 - v_2)^2 = 0.
\end{equation}
We note from (\ref{cs}) that $-\sigma_{\rm III'}(t)/t$ is a Hamiltonian for the
\PIIIprime system, so we can introduce the corresponding 
$\tau$-function by
\begin{equation}\label{cs1a}
\sigma_{\rm III'}(t) = -t {d \over dt} \log \tau_{\rm III'}(t).
\end{equation}
Now the equation (\ref{cs1}) with $v_1 = v_2 = a$ is identical to
(\ref{csp1}), so comparison of (\ref{cs1a}) and (\ref{csp}) gives
the $\tau$-function evaluation
\begin{equation}\label{cs1b}
E_1^{\rm bulk}(0;2s) = \tau_{\rm III'}(\pi^2s^2) \Big |_{v_1=v_2=-1/2}.
\end{equation}
The boundary condition satisfied by $ \sigma_{\rm III'}(t) $ is the $ a=-1/2 $ 
case of (\ref{csp2}),
\begin{equation}\label{Obulk-BC}
   \sigma_{\rm III'}(t) \mathop{\sim}\limits_{t \to 0^+}
   \frac{\sqrt{t}}{2\pi} \left[ 1+\frac{\sin 2\sqrt{t}}{2\sqrt{t}}
                         \right]
                   \mathop{\sim}\limits_{t \to 0^+}
   \frac{\sqrt{t}}{\pi} .
\end{equation}

\subsection{Cumulative distribution of the largest eigenvalue in
the scaled infinite GOE}
The GOE has the property that
to leading order the support of the spectrum is confined
to the interval $[-\sqrt{2N}, \sqrt{2N}]$. It was shown in
\cite{Fo93a} that by scaling the eigenvalues
\begin{equation}\label{1}
\lambda \mapsto \sqrt{2N} + {\lambda \over \sqrt{2} N^{1/6}},
\end{equation}
so that the origin is at the right hand edge of the leading support and
the eigenvalue positions then measured in units of $1/ \sqrt{2} N^{1/6}$, the
distribution functions describing the eigenvalues in the neighbourhood
of this edge (referred to as a soft edge since the density on both sides
is non-zero) are well defined.

It was shown by Tracy and Widom \cite{TW96} (see \cite{Fo99b} for a
simplified derivation) that
\begin{align}\label{2}
  E_1^{\rm soft}(0;(s,\infty)) := F_1(s) 
  & := \lim_{N \to \infty} 
       E_1\Big (0;(\sqrt{2N} + {s \over \sqrt{2} N^{1/6}}, \infty); N \Big )
  \nonumber \\
  & = e^{- {1 \over 2} \int_s^\infty (t-s) q^2(t) \, dt}
      e^{{1 \over 2} \int_s^\infty q(t) \, dt}
\end{align}
where $q(t)$ is the solution of the non-linear equation
\begin{equation}\label{3}
q'' = tq + 2q^3,
\end{equation}
subject to the boundary condition
\begin{equation}\label{4}
q(t) \sim - {\rm Ai}(t) \qquad {\rm as} \qquad t \to \infty,
\end{equation}
where Ai$(t)$ denotes the Airy function.
(Here we have replaced $q$ by $-q$ relative to its use in the original work;
this is valid because (\ref{3}) is unchanged by this mapping.)
Since the general Painlev\'e II equation reads
\begin{equation}\label{5}
q'' = t q + 2q^3 + \alpha,
\end{equation}
(\ref{3}) is the special case $\alpha = 0$ of \PII. Thus (\ref{2}) represents
an explicit evaluation of the gap probability in terms of a
Painlev\'e transcendent. It is the objective of this subsection to show that
in fact  (\ref{2}) can be identified as a $\tau$-function
corresponding to the Painlev\'e II system with $\alpha = 0$.
Consequently its logarithmic derivative satisfies a single nonlinear
differential equation.

Now, in the case of the probability analogous to $F_1(s)$
in the infinite, scaled Gaussian unitary ensemble (GUE), the known exact
evaluation \cite{TW94a} allows one to immediately  make an identification
with a $\tau$-function \cite{FW00}. It is relevant for the purpose of
identifying (\ref{2}) to revise the theory underlying this result.
Tracy and Widom \cite{TW94a} have derived the result
\begin{equation}\label{6}
  E_2^{\rm soft}(0;(s,\infty)) := F_2(s) 
  := \lim_{N \to \infty} E_2(0;(\sqrt{2N} + s/\sqrt{2}N^{1/6}, \infty);N)
   = \exp \Big ( - \int_s^\infty R(t) \, dt \Big ),
\end{equation}
where $R(t)$ satisfies the second order second degree differential equation
\begin{equation}\label{7}
(R'')^2 + 4R'((R')^2 - t R' + R) = 0,
\end{equation}
and have furthermore derived the alternative formula
\begin{equation}\label{7'}
F_2(s) = e^{- \int_s^\infty (t - s) q^2(t) \, dt},
\end{equation}
where $q(t)$ is the same Painlev\'e II transcendent as in (\ref{2}).

To see how the evaluation (\ref{6}) relates to a $\tau$-function for
the Painlev\'e II system, we recall that
in the Hamiltonian formalism of the \PII equation \cite{Ok86}, one defines
a  Hamiltonian $H_{\rm II}$ by
\begin{equation}\label{8}
H_{\rm II} = - {1 \over 2} (2 q^2 - p + t) p - (\alpha + {1 \over 2}) q.
\end{equation}
The canonical coordinate $q$ and momenta $p$ must satisfy the Hamilton
equations (\ref{6.1b}).
Elimination of the variable $p$ between these 
equations  shows that $q$ satisfies
the Painlev\'e II equation (\ref{5}). Furthermore the Hamiltonian
(\ref{8}), regarded as a function of $t$, satisfies the second order
second degree differential equation
\begin{equation}\label{10}
  (H_{\rm II}'')^2 + 4(H_{\rm II}')^3 + 2 H_{\rm II}'(t H_{\rm II}' - H_{\rm II})
   - {1 \over 4} (\alpha + {1 \over 2})^2 = 0,
\end{equation}
referred to as the Jimbo-Miwa-Okamoto $\sigma$ form for \PII. It is also
straightforward to show that $H$ can be expressed in terms of the Painlev\'e
II transcendent $q$ according to
\begin{equation}\label{10'}
H_{\rm II} = {1 \over 2} (q')^2 - {1 \over 2}(q^2 + {1 \over 2}t)^2 -
(\alpha + {1 \over 2}) q.
\end{equation}
Finally, we recall that the $\tau$-function associated with the Painlev\'e
II Hamiltonian is defined by
\begin{equation}\label{10''}
H_{\rm II} = {d \over dt} \log \tau_{\rm II}.
\end{equation}

Setting
\begin{equation}\label{2.44}
u(t; \alpha +  {1 \over 2}) = - 2^{1/3} H_{\rm II}(-2^{1/3} t)
\end{equation}
we see from (\ref{10}) that $u$ satisfies the equation
\begin{equation}\label{11}
  (u'')^2 + 4 u'\left[ (u')^2 - tu' + u \right] - (\alpha + {1 \over 2})^2 = 0.
\end{equation}
Comparison of (\ref{11}) with (\ref{7}) shows
\begin{equation}\label{12}
R(t) = u(t;0) = - 2^{1/3} H_{\rm II}(-2^{1/3}t) \Big |_{\alpha = -1/2}.
\end{equation}
In light of this identification, comparison of (\ref{6}) and (\ref{10''})
then shows,
\begin{equation}\label{13}
F_2(s) = \tau_{\rm II}(-2^{1/3} s) \Big |_{\alpha = - 1/2}.
\end{equation}
The appropriate boundary condition for this $\tau$-function is most simply 
expressed in terms of $R(t)$,
\begin{equation}\label{Usoft_BC}
   R(t) \mathop{\sim}\limits_{t \to \infty}
   [{\rm Ai'}(t)]^2 - t[{\rm Ai}(t)]^2 .
\end{equation}
Thus, up to a scale factor, $F_2(s)$ is precisely the
$\tau$-function associated with the Hamiltonian (\ref{8}) for the
Painlev\'e II system with $\alpha = -1/2$.
A curious feature of (\ref{12}), which follows from (\ref{10'}), is
that $R(t)$ is naturally expressed in terms of the Painlev\'e II
transcendent $q = q(t; -1/2)$, whereas the result (\ref{7'}) involves
the Painlev\'e II transcendent with $\alpha = 0$. In particular, (\ref{6})
and (\ref{7'}) give
\begin{equation}\label{14}
R'(t) = - q^2(t;0)
\end{equation}
while (\ref{12}), (\ref{8}) and the first of the Hamilton equations
(\ref{6.1b}) give
\begin{equation}\label{15}
  R'(t) = - {1 \over 2^{1/3}} \left[ q'(t,-1/2) + q^2(t,-1/2) + {t \over 2}
                              \right]\Big|_{t \mapsto - 2^{1/3} t}.
\end{equation}
In fact, as noted in \cite{FW00}, for $\epsilon = \pm 1$, it is true that
\cite{Gr99}
\begin{equation}\label{16}
- \epsilon 2^{1/3} q^2(-2^{-1/3} t,0) =
{d \over dt} q(t, {1 \over 2} \epsilon) - \epsilon
q^2(t, {1 \over 2} \epsilon)  - {1 \over 2}  \epsilon t,
\end{equation}
which reconciles (\ref{15}) with (\ref{14}).

We are now in a position to identify (\ref{2}) with a $\tau$-function.
The formula (\ref{15}) is just the special case $a=0$ of the identity
\begin{equation}\label{17}
  {d \over dt} H_{\rm II}(t) \Big|_{\alpha = a -1/2}
  = - 2^{-1/3} {d \over dt} u(-2^{-1/3} t;a) 
  = - {1 \over 2} \left[ q'(t,a-1/2) + q^2(t,a-1/2) + {t \over 2} \right] ,
\end{equation}
which is  derived from (\ref{2.44}), (\ref{8}) and the first of the Hamilton
equations (\ref{6.1b}). 
To make use of this result we first note
that the equation (\ref{2}) can be written
\begin{equation}
F_1(s) = e^{- {1 \over 2} \int_s^\infty (t-s)(q^2(t) + q'(t)) \, dt}.
\end{equation}
The identity (\ref{17}) with $a=1/2$ allows this in turn to be rewritten as
\begin{align}\label{18}
  F_1(s) 
  & = \exp \bigg( -\int_s^\infty (t-s) {d \over dt}
      \Big[ 2^{-1/3} u(-2^{-1/3}t;1/2) - {t^2 \over 8} \Big] \, dt \bigg)
  \nonumber \\
  & = \exp \bigg( \int_s^\infty 
      \Big[ 2^{-1/3} u(-2^{-1/3}t;1/2) - {t^2 \over 8} \Big] \, dt \bigg)
  \nonumber \\
  & = \exp \bigg( -\int_s^\infty
      \Big[ H_{\rm II}(t) \Big|_{\alpha = 0} + {t^2 \over 8} \Big] \, dt \bigg) ,
\end{align}
where the final equality follows from (\ref{2.44}). 

We now
associate with $H$ the auxiliary Hamiltonian
\begin{equation}\label{2.54}
h_{\rm II} = H_{\rm II} + {t^2 \over 8}.
\end{equation}
Of course, the Hamilton equations (\ref{6.1b}) remain valid for $H$
replaced by $h$, so $h$ is also a Hamiltonian for the same Painlev\'e II
system.  Introducing the corresponding $\tau$-function by
\begin{equation}\label{25a}
h_{\rm II} = {d \over dt} \log \tau_{h_{\rm II}},
\end{equation}
we see from (\ref{25a}) that
\begin{equation}\label{19}
F_1(s) = \tau_{h_{\rm II}}(s) \Big |_{\alpha = 0},
\end{equation}
which is our sought result. Note that $h_{\rm II}$ satisfies (\ref{10}) with the
substitution
$H_{\rm II} = h_{\rm II} - t^2/8$. It follows from (\ref{10'}), (\ref{2.54})
and (\ref{4}) that we
seek the solution of this equation with $\alpha = 0$ and such that 
\begin{equation}\label{bc19}
  h_{\rm II}(t) \mathop{\sim}\limits_{t \to \infty}
   \frac{1}{2}{\rm Ai}(t) 
   + \frac{1}{2}\left\{ [{\rm Ai}'(t)]^2 - t[{\rm Ai}(t)]^2 \right\} .
\end{equation}

Unlike the situation with $E_1^{\rm bulk}(0;2s)$, in which the corresponding
finite system gap probability $E_1(0;(-s,s);e^{-x^2/2};N)$ is itself a
$\tau$-function, there is no known Painlev\'e transcendent evaluation of
the finite $N$ quantity in the definition (\ref{2}) of $F_1(s)$.
Nonetheless, 
 (\ref{19}) can be obtained as a limiting sequence of finite $N$
Painlev\'e transcendent evaluations, 
which in fact is how we were led to
(\ref{19}) in the first place
\cite{FW01a}. The finite $N$ results are not for
gap probabilities though\footnote{Since completing this work the gap
probability $E_1(0;(s,\infty);e^{-x/2};N)$ has been evaluated as a
\PV $\tau$-function \cite{FW02}, and it scales to $F_1(s)$.}. 
Rather they relate to the quantity
$f_{Nl}^{(\rm inv)}$ specifying the number of fixed point free involutions
of $\{1,2,\dots,2N\}$ constrained so that the length of the maximum decreasing
subsequence is less than or equal to $2l$. This is specified by the
generating function
\begin{equation}
P_l(t) := e^{-t^2/2} \sum_{N=0}^\infty
{t^{2N} \over 2^{2N} N!} {f_{Nl}^{(\rm inv)} \over (2N-1)!!},
\end{equation}
which from the work of Rains \cite{Ra98} (see also
\cite{BF00}) has the integral representation
\begin{equation}
P_l(t) = {e^{-t^2/2} \over l!}
\Big ( {1 \over 2 \pi} \Big )^l
\int_0^\pi d \theta_1 \cdots \int_0^\pi d \theta_l \,
e^{2t \sum_{j=1}^l \cos \theta_j} \prod_{j=1}^l |1 - z_j|^2
\prod_{1 \le j < k \le l} |1 - z_j z_k|^2 |z_j - z_k|^2,
\end{equation}
where $z_j := e^{i \theta_j}$. Although not at all obvious
from the definition, it has been proved in
\cite{BR99b} that
\begin{equation}
  \lim_{l \to \infty} P_l\big({1 \over 2}(l - s (l/2)^{1/3})\big)
   = F_1(s).
\end{equation}
The significance of this result from the present perspective
is that we have recently shown \cite{FW01a} 
$P_l(t)$ to be equal to the $\tau$-function for a certain Painlev\'e V
system which scales to the result (\ref{19}) (the evaluation
of $P_l(t)$ in terms of a  transcendent  related to
Painlev\'e V  was first given by Adler and van Moerbeke
\cite{AV99b}).

\subsection{Cumulative distribution of the smallest eigenvalue in the
scaled infinite LOE}
In the LOE, as $N \to \infty$ the spacing between the eigenvalues in
the neighbourhood of the origin (referred to as the hard edge because
the eigenvalue density is strictly zero for $x < 0$) is of order
$1/N$. With the scaling
\begin{equation*}
\lambda \mapsto {\lambda \over 4 N},
\end{equation*}
the distribution functions describing the eigenvalues near the hard edge
have well defined limits \cite{Fo93a}. Our interest is in
\begin{equation*}
E_1^{\rm hard}(0;(0,s);(a-1)/2) :=
\lim_{N \to \infty} E_1(0;(0,{s \over 4N}); x^{(a-1)/2} e^{-x/2};N),
\end{equation*}
which is equal to the probability of no eigenvalues in the interval
$(0,s)$ of the scaled, infinite LOE, or equivalently to the cumulative
distribution of the smallest eigenvalue in the ensemble. It has been
shown to have the Painlev\'e transcendent evaluation \cite{Fo99b}
\begin{equation}\label{30}
  E_1^{\rm hard}(0;(0,s);(a-1)/2)  =
 \exp \Big( - {1 \over 8} \int_0^s \Big( \log {s \over t} \Big)
               q^2(t) \, dt \Big)
 \exp \Big( - {1 \over 4} \int_0^s {q(t) \over \sqrt{t}} \, dt \Big),
\end{equation}
where $q(t)$ satisfies the nonlinear equation
\begin{equation}\label{qt}
t(q^2 - 1) (tq')' = q(tq')^2 + {1 \over 4}(t-a^2) q +
{1 \over 4} t q^3 (q^2 - 2).
\end{equation}
This equation, which is to be solved subject to the boundary condition
\begin{equation}\label{qs}
  q(t) \mathop{\sim}\limits_{t \to 0^+} J_a(\sqrt{t})
  \mathop{\sim}\limits_{t \to 0^+} {1 \over 2^a \Gamma(1+a)} t^{a/2},
\end{equation}
is transformed \cite{TW94b} via the substitutions
\begin{equation}\label{dm1}
t=x^2, \quad q(t) = {1 + y(x) \over 1 - y(x)} 
\end{equation}
to the \PV equation (\ref{6.1c}) for $y(x)$ with parameters
\begin{equation}\label{2.64'}
\alpha = {a^2 \over 8}, \quad \beta = - {a^2 \over 8}, \quad
\gamma = 0, \quad \delta = -2.
\end{equation}
In this subsection we will show that (\ref{30}) can be identified with a
$\tau$-function corresponding to the \PV Hamiltonian (\ref{6.1}).

To begin we observe that
\begin{equation*}
\int_0^s {q(t) \over \sqrt{t}} \, dt =
\int_0^s \Big ( \log s - \log t \Big )
{d \over dt} \Big (\sqrt{t} q(t)  \Big ) \, dt,
\end{equation*}
in which use is made of (\ref{qs}) for its derivation. Hence we can write
\begin{align}\label{stb}
  E_1^{\rm hard}(0;(0,s);(a-1)/2)
  & = \exp \Big( - {1 \over 8} \int_0^s( \log s - \log t )
                 \left[ q^2 + t^{-1/2} q + 2 t^{1/2} q' \right] \, dt \Big)
  \nonumber \\
  & = \exp \Big( - {1 \over 4} \int_0^{\sqrt{s}} ( \log s - \log t )
                 \left[ x {d q \over dx}  + q + x q^2 \right] \, dx \Big).
\end{align}
But it follows from (\ref{dm1}) that
\begin{equation*}
{d q \over dx} = {2 \over (1-y)^2} {dy \over dx},
\end{equation*}
and thus
\begin{equation}\label{stf}
x {dq \over dx} + q + x q^2 =
{1 \over (1-y)^2}(2x {dy \over dx} - y^2 + 4xy + 1) + x.
\end{equation}

Consider now the Hamiltonian (\ref{6.1}). With the replacements
\begin{equation}\label{2.66e}
q \mapsto y, \quad p \mapsto z, \quad t \mapsto \eta x, \quad
H_{\rm V} \mapsto {1 \over \eta} \tilde{H}_{\rm V}
\end{equation}
it reads
\begin{multline}\label{2.66'}
  x\tilde{H}_{\rm V}(y,z) = y(y-1)^2z^2
    - \Big \{ (v_2 - v_1)(y-1)^2 - 2(v_1+v_2)y(y-1) +\eta xy \Big \}z \\
    + (v_3 - v_1)(v_4 - v_1)(y-1)
\end{multline}
According to (\ref{2.10'}), the remark below (\ref{2.10'}) in
parenthesis  and (\ref{2.8'}), the parameter values (\ref{2.64'})
correspond to the Hamiltonian (\ref{2.66'}) with
\begin{equation}\label{2.66a}
\eta = 2, \quad
v_1 = - v_3 = - {1 \over 4} (a-1), \quad
v_2 = - v_4 = {1 \over 4} (a+1).
\end{equation} 
Furthermore we need to add a term $ -\frac{1}{4}(a^2-1) $ to the Hamiltonian
in order that a well-defined limit for the auxiliary Hamiltonian specified
below exists as $ x \to 0^+ $.
Making use of the Hamilton equations it follows that with these parameter
values 
\begin{align}
  {d \over dx} (x \tilde{H}_{\rm V}) 
  & = - 2 yz \label{stq} \\
  x {d y \over dx} 
  & = 2 y (y-1)^2 z - {a \over 2} (y-1)^2 + y(y-1) - 2xy. \label{stq'}
\end{align}
Substituting for $yz$ in (\ref{stq'}) using (\ref{stq}) we see that
\begin{equation*}
{1 \over (1-y)^2} \left[ 2x {dy \over dx} - y^2 + 4xy + 1 \right] =
-2 \left[ {d \over dx} (x \tilde{H}_{\rm V}) + {a - 1 \over 2} \right] .
\end{equation*}
Substituting this in (\ref{stf}) then gives
\begin{equation}\label{stg}
x {dq \over dx} + q + x q^2 = - 2
\left[ {a-1 \over 2} - {1 \over 2} x + {d \over dx} (x \tilde{H}_{\rm V}) \right] .
\end{equation}
Finally, substituting (\ref{stg}) in (\ref{stb}) and integrating
by parts we arrive at the result
\begin{equation}
  E_1^{\rm hard}(0;(0,s);(a-1)/2)) =
  \exp \int_0^{\sqrt{s}} \Big( {a-1 \over 2} - {1 \over 4} x + \tilde{H}_{\rm V}
                         \Big) \, dx.
\end{equation}
Thus if we define the auxiliary Hamiltonian and corresponding
$\tau$-function for the \PV system by
\begin{equation}\label{hV-defn}
\tilde{h}_{\rm V} = \tilde{H}_{\rm V} - {1 \over 4} x + {a-1 \over 2}, \qquad
\tilde{h}_{\rm V} = {d \over dx} \log \tau_{\tilde{h}_{\rm V}}
\end{equation}
we obtain the sought $\tau$-function evaluation
\begin{equation}\label{2.71}
E_1^{\rm hard}(0;(0,s);(a-1)/2) = \tau_{\tilde{h}_{\rm V}}(\sqrt{s}) \Big |_{
{\scriptstyle \eta = 2 \atop \scriptstyle 
v_1 = - v_3 = - {1 \over 4} (a-1)} \atop  \scriptstyle
v_2 = - v_4 = {1 \over 4} (a + 1)}.
\end{equation}
Note that with the parameters (\ref{2.66a}) it follows from (\ref{2.10a})
and (\ref{2.66e}) that
\begin{equation}\label{no}
  x \tilde{h}_{\rm V} = \sigma_{\rm V}(x) - {1 \over 4} x^2
    + {(a-1) \over 2} x - {a(a-1) \over 4},
\end{equation}
where $\sigma_{\rm V}(x)$ satisfies (\ref{2.g}) with $t \mapsto 2x$.
The boundary condition for this Hamiltonian is
\begin{equation}\label{Ohard-BC}
  x\tilde{h}_{\rm V}(x) \mathop{\sim}\limits_{x \to 0^+}
  - \frac{1}{2}xJ_a(x)
  - \frac{1}{4}x^2 \left[ J^2_a(x) - J_{a+1}(x)J_{a-1}(x) \right] .
\end{equation}

The $\tau$-function evaluation of $E_1^{\rm hard}$ differs from those of
$E_1^{\rm bulk}$ and $E_1^{\rm soft}$ in that no finite $N$ quantity
is known which is itself a $\tau$-function, has an interpretation 
as a probability, and which scales to
(\ref{2.71}). 

\section{Symplectic matrix ensembles} 
\setcounter{equation}{0}
\subsection{Finite $N$ ensembles}
With $N$ finite there is in fact only one symplectic  matrix ensemble ---
the circular symplectic ensemble ---
for which the gap probability can be written in terms of
Painlev\'e transcendents using results known in the literature\footnote{
Since completing this work the gap probability
$E_4(0;(s,\infty);e^{-s};N)$ has been evaluated as the arithmetic mean of
two \PV $\tau$-functions, the Hamiltonians of which satisfy the same
differential equation.}.
With $E_\beta(0;(-\phi,\phi);N)$ denoting the probability that there are no
eigenvalues in an interval $(-\phi,\phi)$ of the circular ensemble specified
by the PDF (\ref{1.6}), this is possible due to the
inter-relationships between gap probabilities due to
Dyson and Mehta \cite{DM63}
\begin{equation}\label{Bs}
  E_4(0;(-\phi,\phi);N) = {1 \over 2} \left\{ 
  E_1(0;(-\phi,\phi);2N)
     + {E_2(0;(-\phi, \phi);2N) \over E_1(0;(-\phi,\phi);2N)} \right\} ,
\end{equation}
implying the evaluation of $E_4$ from knowledge of the evaluation of
$E_1$ and $E_2$

Regarding the latter, let $\phi$ be related to $s$ via the
stereographic projection formula (\ref{1.6a}) with $\theta \mapsto \phi$,
$x \mapsto s$. Then from the relationship between the circular ensemble
and Cauchy ensemble we have
\begin{align*}
  E_1(0;(-\phi,\phi);2N) 
  & = E_1(0;(-s,s); (1+x^2)^{-(N+1/2)}; 2N) \nonumber \\
  E_2(0;(-\phi, \phi);2N) 
  & = E_2(0;(-s,s); (1+x^2)^{-2N}; 2N) \label{3.1'}.
\end{align*}
Now we know from (\ref{F1}) that
\begin{equation*}
E_1(0;(-s,s);
(1+x^2)^{-(2N+1)/2}; 2N) = E_2(0;(0,s^2);x^{-1/2}(1+x)^{-2N};N)
\end{equation*}
while an identity in \cite{Fo99a} gives
\begin{multline}\label{Bs1}
 E_2(0;(-s,s);(1+x^2)^{-2N}; 2N) \\
 =  E_2(0;(0,s^2);x^{-1/2}(1+x)^{-2N};N) E_2(0;(0,s^2);x^{1/2}(1+x)^{-2N};N).
\end{multline}
Thus we have
\begin{align}\label{3.3}
  E_4(0;(-\phi,\phi);N) 
  & = {1 \over 2} \Big\{ E_2(0;(0,s^2); x^{-1/2}(1+x)^{-2N};N)
         + E_2(0;(0,s^2); x^{1/2}(1+x)^{-2N};N) \Big\}
  \nonumber \\
  & = {1 \over 2} \Big\{ E_2(0;(-1,(s^2-1)/(s^2+1));(1+x)^{-1/2}(1-x)^{1/2};N)
  \nonumber \\
  & \qquad + E_2(0;(-1,(s^2-1)/(s^2+1));(1+x)^{1/2}(1-x)^{-1/2};N)
                  \Big\}
  \nonumber \\
  & = {1 \over 2}  \left\{
       \tau_{\tilde{h}_{\rm VI}} \Big( {s^2 \over s^2 + 1} \Big)
            \Big |_{b_1 = b_2 = N  \atop b_3=0, \: b_4 = 1/2}
     + \tau_{\tilde{h}_{\rm VI}} \Big( {s^2 \over s^2 + 1} \Big)
            \Big |_{b_1 = b_2 = N \atop b_3=0, \: b_4 = -1/2} \right\}
\end{align}
where the final equality follows from (\ref{2.sfj}). 
Recalling that
$\tilde{h}_{\rm VI}$ is defined as the RHS of (\ref{h7}), we see from the
fact that $h_{\rm VI}$ satisfies (\ref{h6}) that both cases of
$\tilde{h}_{\rm VI}$ in (\ref{3.3}) satisfy the same differential equation.
Comparing (\ref{Bs}) and (\ref{3.3}) shows the $\tau$-functions in the latter
also give the orthogonal and unitary symmetry gap probabilities,
\begin{equation}\label{3.3a}
E_1(0;(-\phi,\phi);2N) =
\tau_{\tilde{h}_{\rm VI}}
\Big ({s^2 \over s^2 + 1} \Big )
  \Big |_{b_1 = b_2 = N \atop
          b_3 = 0, \: b_4 =  1/2}
\end{equation}
(which is equivalent to a special case of (\ref{au2})) and
\begin{equation}\label{3.3b}
E_2(0;(-\phi,\phi);2N) =
\tau_{\tilde{h}_{\rm VI}}
\Big ({s^2 \over s^2 + 1} \Big )
  \Big |_{b_1 = b_2 = N \atop
          b_3 = 0, \: b_4 =  1/2}\tau_{\tilde{h}_{\rm VI}}
\Big ({s^2 \over s^2 + 1} \Big )
  \Big |_{b_1 = b_2 = N \atop
          b_3 = 0, \: b_4 =  -1/2}.
\end{equation}

\subsection{Bulk gap probability}
In an obvious notation, the bulk scaled limit of (\ref{Bs}) gives the
formula \cite{DM63}
\begin{equation}\label{Bsj}
  E_4^{\rm bulk}(0;s) = {1 \over 2} \left\{ 
  E_1^{\rm bulk}(0;2s) + {E_2^{\rm bulk}(0;2s) \over E_1^{\rm bulk}(0;2s)}
                                    \right\} .
\end{equation}
Using the formula (\ref{csp}) for $E_1^{\rm bulk}$ and a formula for
$E_2^{\rm bulk}$ deduced from the analogue of (\ref{Bs1}), we have
previously shown \cite{Fo99a} that this implies the Painlev\'e
transcendent evaluation
\begin{equation}\label{Bs2}
  E_4^{\rm bulk}(0;s) = {1 \over 2} \left\{
    \exp \Big( - \int_0^{(\pi s)^2}
    \sigma_B(t) \Big|_{a = -1/2} \, {dt \over t}  \Big)
  + \exp \Big( - \int_0^{(\pi s)^2}
    \sigma_B(t) \Big|_{a = 1/2} \, {dt \over t} \Big) \right\}
\end{equation}
where $\sigma_B$ is specified by (\ref{csp1}). Notice that the differential
equation (\ref{csp1}) is unchanged by $a \mapsto -a$, so
$\sigma_B(t) \Big |_{a = -1/2}$ and $\sigma_B(t) \Big |_{a = 1/2}$ differ
in their characterization only by the boundary condition. The
definition (\ref{cs1a}) of $\tau_{\rm III'}(t)$ and the characterization of
$\sigma_{\rm III'}$ therein as the solution of (\ref{cs1}) gives that
(\ref{Bs2}) is equivalent to the $\tau$-function formula
\begin{equation}
  E_4^{\rm bulk}(0;s) = {1 \over 2} \left\{
  \tau_{\rm III'}(\pi^2 s^2) \Big|_{v_1 = v_2 = -1/2} +
  \tau_{\rm III'}(\pi^2 s^2) \Big|_{v_1 = v_2 = 1/2} \right\}.
\end{equation}
The boundary condition for $ \sigma_{\rm III'}(t) $ when $ v_1=v_2=-1/2 $ is given
by (\ref{Obulk-BC}) while the corresponding condition for $ v_1=v_2=1/2 $ is,
from (\ref{csp2}),
\begin{equation}\label{Sbulk-BC}
   \sigma_{\rm III'}(t) \mathop{\sim}\limits_{t \to 0^+}
   \frac{\sqrt{t}}{2\pi} \left[ 1-\frac{\sin 2\sqrt{t}}{2\sqrt{t}}
                         \right]
                   \mathop{\sim}\limits_{t \to 0^+} 
   \frac{t^{3/2}}{3\pi} .
\end{equation}

\subsection{Soft edge scaling}
For the finite Gaussian ensemble the analogue of (\ref{Bs}) is the
coupled equations  \cite{FR99}
\begin{multline}\label{Bs3}
  E_2(0;(s,\infty);e^{-x^2};2N) \\
  = E_1(0;(s,\infty); e^{-x^2/2}; 2N)  \left[
    E_1(0;(s,\infty); e^{-x^2/2}; 2N+1) + E_1(1;(s,\infty); e^{-x^2/2}; 2N+1)
                                         \right] \\
    + E_1(0;(s,\infty); e^{-x^2/2}; 2N+1) E_1(1;(s,\infty); e^{-x^2/2}; 2N)
\end{multline}
\begin{equation}\label{Bs3'}
  E_4(0;(s,\infty);e^{-x^2};N) =
  E_1(0;(s,\infty); e^{-x^2/2}; 2N) + E_1(1;(s,\infty); e^{-x^2/2}; 2N+1)
\end{equation}
Here the only known quantity is $E_2$. In the soft edge scaling limit the 
number of distinct quantities in (\ref{Bs3}) is reduced, and one obtains
the analogue of (\ref{Bs}) \cite{FR99},
\begin{equation}\label{fyf}
  E_4^{\rm soft}(0;(s,\infty)) = {1 \over 2} \left\{
  E_1^{\rm soft}(0;(s,\infty)) +
  {E_2^{\rm soft}(0;(s,\infty)) \over E_1^{\rm soft}(0;(s,\infty))} \right\} .
\end{equation}
As noted in \cite{Fo99b}, it follows from (\ref{2}) and
(\ref{7'}) that
\begin{equation}\label{fy}
  E_4^{\rm soft}(0;(s,\infty)) = {1 \over 2} \Big (
  e^{- {1 \over 2} \int_s^\infty (t-s) q^2(t) \, dt}
  e^{{1 \over 2} \int_s^\infty q(t) \, dt} +
  e^{- {1 \over 2} \int_s^\infty (t-s) q^2(t) \, dt}
  e^{-{1 \over 2} \int_s^\infty q(t) \, dt} \Big ),
\end{equation}
where $q(t)$ satisfies (\ref{3})
(this result was first derived in a direct calculation
\cite{TW96}). The first term in (\ref{fy}) has in (\ref{19})
been identified as a $\tau$-function. The second term differs from the
first only in the sign of $q(t)$. Since the differential equation
(\ref{3}) is unchanged by the replacement $q \mapsto - q$, we see
that we can write the second term in (\ref{fy}) in a form
formally identical to the first. Consequently
\begin{equation}\label{fyt}
  E_4^{\rm soft}(0;(s,\infty)) = {1 \over 2} \left\{
  \tau_{h_{\rm II}}^{(1)}(s) + \tau_{h_{\rm II}}^{(2)}(s) \right\} \Big|_{\alpha = 0}
\end{equation}
where $h$ in $\tau_h^{(1)}(s)$ is as in (\ref{19}), while $h$ in
$\tau_h^{(2)}(s)$ is characterized  as the solution of the same
differential equation as in (\ref{19}), but with the boundary condition
\begin{equation}\label{bc20}
  h_{\rm II}(t) \mathop{\sim}\limits_{t \to \infty}
   - \frac{1}{2}{\rm Ai}(t) 
   + \frac{1}{2}\left\{ [{\rm Ai}'(t)]^2 - t[{\rm Ai}(t)]^2 \right\}
\end{equation}
which results by substituting (\ref{4}) without the minus sign on the RHS
in (\ref{10'}) with $a=1/2$ and recalling (\ref{2.54}) (c.f.~(\ref{bc19})).

\subsection{Hard edge scaling}
In the case of the finite $N$ Laguerre ensemble, the probabilities for the
different symmetry classes of no eigenvalues in the interval $(0,s)$
at the hard edge of the spectrum are related by coupled equations of the
form (\ref{Bs3}), (\ref{Bs3'})
\cite{FR99}. Consequently, in the scaled limit one obtains
the analogue of (\ref{fyf}) \cite{FR99}
\begin{equation}\label{fyft}
  E_4^{\rm hard}(0;(0,s);a+1) =
  {1 \over 2} \left\{ E_1^{\rm hard}(0;(0,s);(a-1)/2) +
  {E_2^{\rm hard}(0;(0,s);a) \over E_1^{\rm hard}(0;(0,s);(a-1)/2)} \right\} ,
\end{equation}
and using (\ref{30}) and the analogous result for $E_2^{\rm hard}$
\cite{TW94b}, we obtain \cite{Fo99b}
\begin{align}\label{ky}
  E_4^{\rm hard}(0;(0,s);a+1) = {1 \over 2} \Bigg\{
  & \exp \Big( - {1 \over 8} \int_0^s \Big( \log{s \over t} \Big) q^2(t)
    \, dt \Big )
  \exp \Big( - {1 \over 4} \int_0^s {q(t) \over \sqrt{t}} \, dt \Big) 
  \nonumber \\
  & \phantom{\Big\{} +
    \exp \Big( - {1 \over 8} \int_0^s \Big( \log{s \over t} \Big) q^2(t)
    \, dt \Big)
    \exp \Big(  {1 \over 4} \int_0^s {q(t) \over \sqrt{t}} \, dt \Big)
    \Bigg\}
\end{align}
where $q(t)$ satisfies (\ref{qt}). As with (\ref{fy}), the first term in 
(\ref{ky}) is the orthogonal ensemble result, which has been
 identified as a $\tau$-function in (\ref{2.71}) above, while
the second term differs from the first only in the sign of $q(t)$.
A further analogy is that (\ref{qt}), like (\ref{3}) is unchanged by the
mapping $q \mapsto - q$, so we have 
\begin{equation}\label{fyt1}
  E_4^{\rm hard}(0;(0,s);a+1) = {1 \over 2} \left\{
  \tau_{\tilde{h}_{\rm V}}^{(1)}(\sqrt{s}) + \tau_{\tilde{h}_{\rm V}}^{(2)}(\sqrt{s})
                                            \right\}
   \Big|_{{\scriptstyle \eta = 2 \atop
           \scriptstyle v_1 = - v_3 = - {1 \over 4}(a-1)} \atop
           \scriptstyle v_2 = - v_4 = {1 \over 4}(a+1)}
\end{equation}
where $\tilde{h}_{\rm V}$ in $\tau_{\tilde{h}_{\rm V}}^{(1)}$ is as in
(\ref{2.71}), while $\tilde{h}_{\rm V}$ in $\tau_{\tilde{h}_{\rm V}}^{(2)}$ satisfies
the same equation (recall (\ref{qt})) but the boundary condition as
determined by (\ref{dm1}), (\ref{2.66'}), (\ref{hV-defn}) and (\ref{qs}) is 
different because a minus sign must now be placed in front of (\ref{qs}). 
Explicitly we have for the second term
\begin{equation}\label{Shard-BC}
  x\tilde{h}_{\rm V}(x) \mathop{\sim}\limits_{x \to 0^+}
    \frac{1}{2}xJ_a(x)
  - \frac{1}{4}x^2 \left[ J^2_a(x) - J_{a+1}(x)J_{a-1}(x) \right]
\end{equation}
(c.f.~(\ref{Ohard-BC})). For each of the bulk, soft and hard edge cases, the
$\tau$-functions in the evaluation of $E_4$ give $E_1$ and $E_2$ according to
formulas analogous to (\ref{3.3a}) and (\ref{3.3b}). These are summarized
in the accompanying table.

\subsection*{Acknowledgement}
This work was supported by the Australian Research Council.
PJF thanks A.~Borodin, P.~Deift and A.R.~Its for stimulating
discussions.


\vfill\eject

%
\thispagestyle{empty}
\renewcommand{\arraystretch}{2.0}
\begin{table}

\begin{sideways}

\small
\begin{tabular}{|c|c|c|c|}

\multicolumn{4}{c}{\rule[-8mm]{0mm}{10mm}
\underline{\large Gap Probabilities in Scaled Random Matrix Ensembles ---
Jardin d'Jimbo-Miwa-Okamoto}} \\

\hline
 Scaling Limit 	&
 Orthogonal     &
 Unitary        &
 Symplectic     \\
\hline

 &
\multicolumn{3}{|c|}{
 \PIIIprime : 
 $ \quad v_1 = v_2 = \pm\half $
 \hskip2cm
 \PV : 
 $ \quad \eta = 1 \quad v_1 = v_2 = v_3 = v_4 = 0 $}	\\
\cline{2-4}

 Bulk		&
\begin{minipage}[c][1.2cm][c]{5cm}{
  \begin{equation*}
     E^{\rm bulk}_1(0;(-s,s)) = 
     \tau_{{\rm III}^{\prime}}(\pi^2s^2)\big|_{-\half}
  \end{equation*}}
\end{minipage} 	&
\begin{minipage}[c][1.9cm][t]{7cm}{
  \begin{align*}
     E^{\rm bulk}_2(0;(-s,s))
     & = \tau_{{\rm III}^{\prime}}(\pi^2s^2)\big|_{-\half}
       \,\tau_{{\rm III}^{\prime}}(\pi^2s^2)\big|_{\half}	\\
     & = \tau_{\rm V}(i\pi s)
  \end{align*}}
\end{minipage} 	&
\begin{minipage}[c][1.2cm][c]{9cm}{
  \begin{equation*}
     E^{\rm bulk}_4(0;(-s/2,s/2)) = 
     \half\tau_{{\rm III}^{\prime}}(\pi^2s^2)\big|_{-\half}
     + \half\tau_{{\rm III}^{\prime}}(\pi^2s^2)\big|_{\half}
  \end{equation*}}
\end{minipage} 	\\
\cline{2-4}

 		&
\multicolumn{3}{|c|}{
\begin{minipage}[c][1.6cm][c]{9cm}{
\begin{equation*}
    \sigma_{{\rm III}^{\prime}}(t)\big|_{v_1=v_2=\mp\half}
    \;\mathop{\sim}\limits_{t \to 0^+}
    \frac{\displaystyle \sqrt{t}}{\displaystyle 2\pi}
    \left[ 1 \pm \frac{\displaystyle \sin 2\sqrt{t}}{\displaystyle 2\sqrt{t}} \right]
\end{equation*}}
\end{minipage}
\hskip0cm
\begin{minipage}[c][1.3cm][c]{2cm}{
  \begin{equation*}
     H_{\rm V}(t) \mathop{\sim}\limits_{t \to 0^+}
      - {t \over \pi}-{t^2 \over \pi^2}
  \end{equation*}}
\end{minipage}} 	\\
\hline

 &
 \multicolumn{3}{|c|}{
 \PII : 
 $ \quad \alpha = 0 $}
      \\
\cline{2-4}

 Soft Edge	&
\begin{minipage}[c][1.2cm][c]{6cm}{
  \begin{equation*}
     E^{\rm soft}_1(0;(s,\infty)) = 
     \tau^{(1)}_{{\rm II}}(s)
  \end{equation*}}
\end{minipage} 	&
\begin{minipage}[c][1.9cm][t]{6cm}{
  \begin{align*}
     E^{\rm soft}_2(0;(s,\infty)) 
     & = \tau^{(1)}_{{\rm II}}(s) \,\tau^{(2)}_{{\rm II}}(s)
     \\
     & = \tau_{{\rm II}}(-2^{1/3}s)\big|_{\alpha=-1/2}
  \end{align*}}
\end{minipage} 	&
\begin{minipage}[c][1.2cm][c]{8cm}{
  \begin{equation*}
     E^{\rm soft}_4(0;(s,\infty)) = 
       \half\tau^{(1)}_{{\rm II}}(s)
     + \half\tau^{(2)}_{{\rm II}}(s)
  \end{equation*}}
\end{minipage} 	\\
\cline{2-4}

		&
\multicolumn{3}{|c|}{
\begin{minipage}[c][1.3cm][t]{8cm}{
  \begin{equation*}
     h^{(1,2)}_{\rm II}(t) \mathop{\sim}\limits_{t \to \infty}
     \pm \half{\rm Ai}(t)                       
     + \half\left\{ [{\rm Ai}'(t)]^2 - t[{\rm Ai}(t)]^2 \right\}
  \end{equation*}}
\end{minipage}
\hskip1cm
\begin{minipage}[c][1.3cm][t]{7cm}{
  \begin{equation*}
     -2^{1/3}H_{\rm II}( -2^{1/3}t)\big|_{\alpha=-1/2} 
      \mathop{\sim}\limits_{t \to \infty} [{\rm Ai}'(t)]^2 - t[{\rm Ai}(t)]^2
  \end{equation*}}
\end{minipage}}	\\
\hline

 &
 \multicolumn{3}{|c|}{
 \PV : 
 $ \quad \eta = 2 $
 $ \quad v_1 = -v_3 = -\quarter(a-1) $
 $ \quad v_2 = -v_4 =  \quarter(a+1) $
 \hskip2cm
 \PIIIprime : 
 $ \quad v_1 = v_2 = a $} \\
\cline{2-4}

 Hard Edge	&
\begin{minipage}[c][1.2cm][c]{6cm}{
  \begin{equation*}
     E^{\rm hard}_1(0;(0,s);(a-1)/2) = 
     \tau^{(1)}_{\rm V}(\sqrt{s})
  \end{equation*}}
\end{minipage} 	&
\begin{minipage}[c][1.9cm][t]{7cm}{
  \begin{align*}
     E^{\rm hard}_2(0;(0,s);a)
     & = \tau^{(1)}_{\rm V}(\sqrt{s})\,\tau^{(2)}_{\rm V}(\sqrt{s}) \\
     & = \tau_{\rm III^{\prime}}(s/4)
  \end{align*}}
\end{minipage} 	&
\begin{minipage}[c][1.2cm][c]{9cm}{
  \begin{equation*}
     E^{\rm hard}_4(0;(0,s);a+1) = 
     \half\tau^{(1)}_{\rm V}(\sqrt{s})
     + \half\tau^{(2)}_{\rm V}(\sqrt{s})
  \end{equation*}}
\end{minipage} 	\\
\cline{2-4}

		&
\multicolumn{3}{|c|}{
\begin{minipage}[c][1.3cm][t]{9cm}{
  \begin{equation*}
     th^{(1,2)}_{\rm V}(t) \mathop{\sim}\limits_{t \to 0^+}
     \mp  \half tJ_a(t)
     - \quarter t^2 \left[ J^2_a(t) - J_{a+1}(t)J_{a-1}(t) \right]
  \end{equation*}}
\end{minipage}
\hskip2cm
\begin{minipage}[c][1.3cm][t]{7cm}{
  \begin{equation*}
     \sigma_{{\rm III}^{\prime}}(t) \mathop{\sim}\limits_{t \to 0^+}
       \quarter t\left[ J^2_a(\sqrt{t}) - J_{a+1}(\sqrt{t})J_{a-1}(\sqrt{t})
              \right]
  \end{equation*}}
\end{minipage}}	\\
\hline

\end{tabular}
\end{sideways}

\end{table}

\end{document}